\documentstyle[psfig]{l-aa}
\def\spose#1{\hbox to 0pt{#1\hss}}
\def\simlt{\mathrel{\spose{\lower 3pt\hbox{$\mathchar"218$}}
     \raise 2.0pt\hbox{$\mathchar"13C$}}}
\def\simgt{\mathrel{\spose{\lower 3pt\hbox{$\mathchar"218$}}
     \raise 2.0pt\hbox{$\mathchar"13E$}}}

\def\msun{{\,M_\odot}}
\def\rs{R_{\rm s}}

\def\tc{t_{\rm c}}

\begin{document}
\thesaurus{11.03.5; 11.09.3; 11.05.2; 11.01.1}
\title{The Influence of Stellar Energetics and Dark Matter on the 
      Chemical Evolution of Dwarf Irregulars} 

\author{F. Bradamante$^{1,2}$, F. Matteucci$^{1,2}$ and A. D'Ercole$^{3}$}
\offprints{F. Matteucci}
\institute{
Dipartimento di Astronomia, Universit\`a di Trieste, Via G.B. Tiepolo 11,
34131 Trieste, Italy \and
SISSA/ISAS, Via Beirut 2-4, 34014 Trieste, Italy \and
Osservatorio Astronomico di Bologna, Via Zamboni 33, 44100 Bologna, Italy}
\maketitle
\markboth{}{}
\begin{abstract}
A chemical evolution model following the
evolution of the abundances of H, He, C, N, O and Fe
for dwarf irregular and blue compact galaxies is presented.
This model takes into account detailed nucleosynthesis and computes
in detail the rates of
supernovae of type II and I. 
The star formation is assumed to have proceeded in short but intense 
bursts. The novelty relative to previous models is that the
development of a galactic wind is studied in detail by
taking into account the energy injected into the interstellar
medium (ISM) from both supernovae and stellar winds from massive stars
as well as the presence of dark matter halos. 
Both metal enriched and normal winds have been considered. \\
Our main conclusions are: 
{\it i)} a substantial amount of dark matter (from 1 to 50 times
larger than the luminous matter) is required in order to avoid 
the complete destruction of such galaxies during strong starbursts, and 
{\it ii)} the energy injected by stellar winds
and type Ia supernovae into the ISM is 
negligible relative to the total thermal energy, and in particular to
the type II supernovae, which in fact, dominate the energetics during 
starbursts. 
\end{abstract}

\section{Introduction}

Dwarf irregulars galaxies (DIG) and blue compact galaxies (BCG) are
very interesting objects to study galaxy evolution, for they are easier
to model because of their generally small sizes and simple structures.
Their stellar populations appear to be mostly young, their metallicity is
low and their gas content is large. All these features indicate that these 
galaxies are poorly evolved objects, and may have undergone discontinuous 
or gasping star formation activity. 
In particular, BCG are the most luminous among 
dwarf irregular 
galaxies: their very low metal content, large gas fraction and very blue 
colours, suggest that star formation may have proceeded in short and 
intense bursts of activity (Searle et al. 1978). 
On the other hand, DIG seem to show a `gasping' activity of star formation
(Tosi et al. 1992).
In the last few years a great deal of theoretical models for DIG and BCG
appeared in the literature. By means of analytical
models, Matteucci and Chiosi (1983) first studied the effects of 
galactic winds on the chemical evolution of these galaxies, and suggested, 
among other hypotheses, that galactic winds of 
variable intensity could explain the observed spread in the physical 
properties of these galaxies.
In the following years, by means of
numerical models, Matteucci and Tosi (1985), Pilyugin (1992, 1993) and
Marconi et al. (1994), studied the effects of 
galactic winds on the chemical evolution of such galaxies and
confirmed the previous results. In these papers 
the possibility of metal enriched winds was also considered
and favoured relative to the normal galactic winds.
This metal enriched wind hypothesis, in fact, allowed them to explain the
observed He abundances versus metal abundances.
However, none of these studies had taken into account either the energy 
injected 
by supernovae and stellar winds into the ISM or the presence of dark 
matter halos which are of fundamental importance for the development of a 
galactic wind. The galactic winds were just assumed to occur every time
the starburst was active. 
\par
Dark matter in dwarf irregular galaxies seems to be quite important as 
shown by recent data indicating an increasing amount of dark matter with 
decreasing luminosity (Skillman 1996), and therefore it cannot be ignored 
in chemical evolution models. Kumai and Tosa (1992) also explored the 
possibility of dark matter in these galaxies in a very simple way, and
suggested that various amounts of dark matter could explain the observed
spread in the $Z\,\, versus \,\, log(M_{gas}/M_{tot})$ relation.
\par
The energetics of supernovae and stellar winds and their interaction
with the ISM, is a difficult problem, still not quite understood.
The only existing models of chemical evolution taking into account the
stellar energetics relative to the occurrence of galactic winds, are those
developed to study the chemical and photometric evolution of 
elliptical galaxies (Arimoto and Yoshii 1987; Matteucci and Tornamb\'e 
1987; Angeletti and Giannone 1990; Ferrini and Poggianti 1994; 
Bressan et al. 1994; Gibson 1994, 1997).
In these models, some specific assumptions relative to the amount of
energy transferred into the ISM are made. Gibson (1994) has shown that
the energetics from stellar winds is negligible in massive ellipticals
whereas it could be important in smaller ones.
For this reason it seems worthwhile to explore the effects of both supernovae
and stellar winds on the evolution of dwarf irregulars and blue compact
galaxies.
\par
The aim of this paper is to take into account both the stellar energetics 
and the presence of dark matter in modelling the chemical evolution of dwarf 
irregular galaxies.
The development of a galactic wind will therefore be calculated in detail
and the existence of metal enriched or differential winds will be taken
into account.
By differential wind we intend a wind carrying out only some heavy elements,
in particular those produced by type II supernovae which are the predominant
supernovae during a starburst (Marconi et al. 1994) and eject material at 
much higher velocities than normal stellar winds. Therefore, under this 
assumption elements such as nitrogen and helium, which are restored 
by low and intermediate mass stars through 
stellar 
winds, will not leave the star forming region whereas oxygen and the other 
$\alpha$-elements, ejected during type II supernova explosions 
are likely to be ejected outside the region and perhaps outside the galaxy. 
Iron 
will only in 
part leave the star forming region since only a fraction of this element 
is produced by SNe II, whereas the bulk of it comes from SNe Ia. 
It should be again said that the differential wind assumption has proven 
to be the most viable solution to the helium problem in blue compact galaxies 
(see Pilyugin 1993 and Marconi et al. 1994).
\par
The paper is organized as follows: in section 2. we will present the 
observational data concerning dwarf irregulars and blue 
compact galaxies,
in section 3. we will describe the chemical evolution model and the
nucleosynthesis prescriptions, in section 4. we will give our model
results and finally in section 5. we will draw some conclusions.

\section{Data sample}

The data we refer to are those presented in Marconi et al. (1994) (mainly 
Pagel et al. 1992 and Campbell 1992) and in Skillmann (1996) (mainly 
Thuan 1995 and Garnett 1990).
\par
Thuan at al. (1995) presented high quality spectrophotometric observations
of 15 supergiant HII regions in 14 BCG. We compare our model results with
their data relative to N, O, and Fe. Their data showed that none of 
the ratios N/O, Fe/O, Ne/O, Ar/O, S/O, depends on the oxygen abundance
indicating that abundance ratios in these galaxies show the pollution
only from massive stars and that all the elements, including nitrogen, 
should have a primary origin (see paragraph 3.1). 
\par
Garnett et al. (1995a) presented UV observations of 7 HII regions in 
low-luminosity dwarf irregular galaxies and the Magellanic Clouds (HST data).
This allowed them to measure the carbon abundances and discuss the evolution
of the C/O ratio in these galaxies. They found that the C/O ratio increases
continuously with increasing O/H and suggested that this can be due to the 
fact that in the most metal poor galaxies, massive star nucleosynthesis 
predominates (oxygen production), while in more metallic objects the delayed 
contribution of intermediate mass stars to carbon is more evident.
However, their comparison with chemical evolution models is not really 
appropriate since they have adopted, for such comparison, models relative 
to the solar neighbourhood, where the star formation mechanism and evolution 
are definitively different from those of dwarf irregular galaxies.

\section{Theoretical Prescriptions}

To calculate the chemical evolution of the star forming region of DIG and 
BCG, we used an improved version of the model 
presented in Marconi et al. (1994).
The main features of our chemical evolution model are the following:
\begin{enumerate}
\item
 one-zone, with instantaneous and complete mixing of gas inside this zone;
\item
 no instantaneous recycling approximation, i.e. the stellar
lifetimes are taken into account;
\item
 the evolution of several chemical elements (He, C, N, O, Fe) due
to stellar nucleosynthesis, stellar mass ejection, galactic wind
powered by supernovae and stellar wind and infall of primordial gas, is 
followed in details.
\end{enumerate}
If $G_i$ is the fractional mass of the element {\it i} in the gas, 
its evolution is
given by the equations:
\begin{eqnarray}
  \dot G_{i}= -\psi(t) X_{i}(t)+R_{i}(t)+
   (\dot G_{i})_{inf} - \dot G_{iw}(t) 
\label{uno}
\end{eqnarray}
where $G_{i}(t)= M_{g}(t)X_{i}(t)/ M_{L}(t_{G})$ is the gas mass
in the form of an element {\it i} normalized to a total luminous 
mass $M_{L}= 10^{9}M_{\odot}$.\\
The quantity $X_{i}(t)=G_{i}(t)/G(t)$  represents 
the abundance by mass of an element {\it i}
and by definition the summation over all the elements present in the gas
mixture is equal to unity.
The quantity ${G(t)}=M_{g}(t)/M_{L}$ is the total fractional mass of 
gas. 
\par
The star formation rate we assume during a burst, $\psi(t)$, is defined as:
\begin{eqnarray}
 \psi(t)\,=\Gamma G(t) 
\end{eqnarray}
where $\Gamma$ is the star formation efficiency (expressed in units of  
$Gyr^{-1}$), and represents the inverse of the timescale of star
formation, namely the timescale necessary to consume all the gas in the star
forming region. More precisely, $\Gamma$ can be expressed as $\nu \eta(t)$
where $\nu$ is the real star formation efficiency, and $\eta(t)$ is a
parameter taking into account the stochastic nature of the 
star formation processes (Gerola et al. 1980) as shown in Matteucci 
and Chiosi (1983), where a detailed description can be found. 
\par
The rate of gas loss via galactic winds for each element is assumed to be
simply proportional to the amount of gas present at the time $t$:
\begin{eqnarray}
 \dot G_{iw}\,=\,w_{i} \, G(t) \, X_{iw}(t) 
\end{eqnarray}
where $X_{iw}(t)=X_{i}(t)$ is the abundance of the element {\it i} in the wind
and in the interstellar medium (ISM);
$w_{i}$ is a free parameter describing the efficiency of the galactic
wind and expressed in $Gyr^{-1}$.
We have considered  normal and differential winds: in the first
case, all the elements are lost at the same efficiency ($w_{i}=w$). 
In the second case, the value of $w_{i}$ has been assumed to be different 
for different elements: in particular, the assumption has been made that 
only the elements produced by type II SNe (mostly $\alpha$-elements and 
one third of the iron) can escape the star forming region. We have made 
this choice following the 
conclusions of Marconi et al. (1994), who showed that models with 
differential winds can better explain the observational constraints of blue 
compact galaxies in general. 
The conditions for the onset of the galactic wind are studied in detail
and described in paragraph $3.2$.
\par
The chemical evolution equations include also an accretion term:
\begin{eqnarray}
 (\dot G_{i})_{inf} \, = \, C \, {(X_{i})_{inf}e^{-(t/\tau)} \over M_L} 
\end{eqnarray}
where $(X_{i})_{inf}$ is the abundance of the element {\it i} in the infalling
gas, assumed to be primordial, $\tau$ is the time scale of mass accretion 
$t_{G}$ is the galactic lifetime, and $C$ is a constant obtained by imposing
to reproduce $M_L$ at the present time $t_{G}$.
The parameter $\tau$ has been assumed to be the same for all dwarf 
irregulars and short enough to avoid unlikely high infall rates at
the present time ($\tau=0.5\cdot 10^{9}$ years). 
\par
Finally, the initial mass function (IMF) by mass, $\phi(m)$, is 
expressed as a power law:
\begin{eqnarray}
  \phi(m) = A \,  m^{-(1+x)}  
\end{eqnarray}
We considered two cases for the IMF:
\begin{enumerate}
\item
 the exponent $x=1.35$ over the mass range $(0.1 \div 100) \,M_{\odot}$
(Salpeter 1955 IMF), and
\item
  $x=1.35$ over the mass range
$(0.1 \div 2) \,M_{\odot}$, and  $x=1.70$ over the mass range $(2 \div 100)
M_{\odot}$ (Scalo 1986 IMF).
\end{enumerate}
\noindent
$A$ is the normalisation constant, obtained with the following condition:
\begin{eqnarray}
   A \int_{m_{L}}^{m_{U}}  m^{-x} \, dm = 1 
\end{eqnarray}

\subsection{Nucleosynthesis Prescriptions} 

The $R_{i}(t)$ term of the equation (1) represents the stellar 
contribution to the enrichment of the ISM, i.e. the rate at which the element 
{\it i} is restored into the ISM from a stellar generation 
(see Marconi et al. 1994). This term contain all the nucleosynthesis
prescriptions:
\begin{enumerate}
\item
for low and intermediate mass stars ($0.8 M_{\odot}\leq M \leq
M_{up}$) we have used Renzini and Voli's (1981) nucleosynthesis
calculations for a value of the mass loss parameter $\eta = 0.33$
(Reimers 1975), and the mixing length $\alpha_{RV} = 1.5$. The standard 
value for $M_{up}$ is 8 $M_{\odot}$; 
\item
 for massive stars ($M>8M_{\odot}$) we have used Woosley's (1987)
nucleosynthesis computations but adopting the relationship between the
initial mass $M$ and the He-core mass $M_{He}$, from
Maeder (1989). It is worth noting that the adopted $M(M_{He}$) relationship
does not substantially differ from the original relationship given by 
Arnett (1978) and from the new one by Maeder (1992) based on models with 
overshooting and $Z=0.001$. These new models show instead 
a very different behaviour of $M(M_{He}$) for stars more massive than 
25 $M_{\odot}$ and $Z=0.02$, but our galaxies never reach 
such a high metallicity;
\item
for the explosive nucleosynthesis products, we have adopted the 
prescriptions by Nomoto et al. (1986) and Thielmann et al. (1993), 
model W7, for type Ia SNe, which we assume to originate from C-O white 
dwarfs in binary systems (see again Marconi et al. 1994 for details).
\end{enumerate}
As already said, nitrogen is a key element to
understand the evolution of galaxies with few star forming events since 
it needs relatively long timescales as well as a relatively high
underlying metallicity to be produced.
The reason is that N is believed to be mostly a secondary element
(secondary elements are those synthesized from metals originally
present in the star and not produced {\em in situ},
while primary elements are those synthesized directly from H and He). 
In fact, N is secondary in massive stars, and mostly secondary and probably 
partly primary in low and intermediate mass stars (Renzini and Voli 1981).
However, some doubts exist at the moment on the amount of primary
nitrogen which can be produced in intermediate 
mass stars due to the uncertanties related
to the occurrence of the third {\em dredge-up} in asymptotic giant branch
stars (AGB). In fact, if Bl\"ocker and Schoenberner (1991) calculations are
correct,
the third {\em dredge-up} in massive AGB stars should not occur and therefore
the amount of primary N produced in AGB stars should be
strongly reduced (Renzini, private communication). 
\par
As a consequence, the only way left to produce a reasonable quantity of N 
during a short burst (no longer than 20 $Myr$) is to require that massive 
stars produce a substantial amount of primary nitrogen. This claim was 
already made by Matteucci (1987) in
order to explain the [N/O] abundances in the solar neighbourhood.
Recently, Woosley et al. (1994) have shown that massive stars can
indeed produce primary nitrogen, and Marconi et al. (1994) and Kunth et al.
(1995) have taken into account this possibility.
In this paper we also have considered this possibility and, since the Woosley
et al. data are not yet available we have used the same
parametrization adopted by Marconi et al. (1994) and Kunth et al. (1995).

\subsection{Energetics} 

Our model presents a new formulation of galactic winds
relative to previous published models of this type (Matteucci and Tosi 1985
and Marconi et al. 1994). In particular, we adopted the 
prescriptions developed for elliptical galaxies 
(Matteucci and Tornamb\'e 1987; Gibson 1994, 1997):
 galaxies develop galactic winds when the gas thermal energy
$E_{g}^{th}(t)$, exceeds its binding energy $E_{g}^{b}(t)$, i.e. when:
\begin{eqnarray}
   E_{g}^{th}(t) \geq E_{g}^{b}(t)
\end{eqnarray}
\par
A detailed treatment of the energetics of the ISM
is considered, in order to compute the gas thermal energy:
\begin{eqnarray}
 E_{g}^{th}(t)= E_{SN}^{th}(t) + E_{sw}^{th}(t)
\end{eqnarray}
 we calculated the energy fraction deposited in the gas
by stellar winds from massive stars $E_{sw}^{th}(t)$, and by supernova 
explosions $E_{SN}^{th}(t)$. Here the supernova contribution is given by:
\begin{eqnarray}
   E_{SN}^{th}(t)=E_{SN II}^{th}(t) + E_{SN Ia}^{th}(t) 
\end{eqnarray}
\noindent
where: 
\begin{eqnarray}
   E_{SN II}^{th} = \int_{0}^{t}\epsilon_{SN} R_{SNII}(t')\,dt' 
\end{eqnarray}
\begin{eqnarray}
   E_{SN Ia}^{th} =\int_{0}^{t}\epsilon_{SN} R_{SNIa}(t')\,dt' 
\end{eqnarray}
\begin{eqnarray}
  E_{sw}^{th}(t)=\int_{0}^{t} \int_{12}^{m_{up}} \phi(m) 
    \psi(t') \epsilon_{w} \, dm \, dt'
\end{eqnarray}
$R_{SNII}(t)$ and $R_{SNIa}(t)$ are the rates of supernova
(II and Ia) explosion, $\phi(m)$ is the IMF and $\psi(t)$ is the star
formation rate. The type Ia and II SN rates are calculated according
to Matteucci and Greggio (1986).
The energy injected and effectively thermalized into the ISM from 
supernova explosions ($\epsilon_{SN}$) and stellar winds from 
massive stars ($ \epsilon_{w}$), are given by the following formulas:
\begin{eqnarray}
  \epsilon_{SN}= \eta_{SN} \, E_{0}
\label{f.sn}
\end{eqnarray}
\begin{eqnarray}
   \epsilon_{w}= \eta_{w} \, E_{w}
\label{f.sw}
\end{eqnarray}
where $E_{0}$ is the total energy released by a supernova explosion, $E_{w}$ is
the energy injected into the ISM by a typical massive star 
through stellar winds during all its lifetime, and $\eta_{SN}$ and 
$\eta_{w}$ are the efficiencies of energy 
transfer from supernova and stellar winds into the ISM, respectively.
We adopted a formulation for $\epsilon_{SN}$ and $\epsilon_{w}$ 
which is described in detail in appendix. 

\par
Our formulation refers to an ideal case characterized by an 
uniform ISM, and no interaction with other supernova explosions 
or interstellar clouds. 
When a supernova explodes the stellar material is violently ejected
into the ISM and its expansion is gradually slowed down by the ISM. 
The energy released by the supernova explosion is assumed to be 
$E_{0}=10^{51} \, erg$, while the energy effectively transferred and 
thermalized into the ISM is given by eq. (13) and depends on the assumed
value for $\eta_{SN}$.
The main parameters used to derive
$\eta_{SN}$, as described in the appendix, are: the initial blast wave energy 
$E_{51}=E_{0}/(10^{51}\, erg)$, the interstellar gas 
density $n_{0}$, and the isothermal sound speed in the ISM
$c_{0,6}=c_{0}/(10^{6}cm \, s^{-1})$. 

For typical values of these parameters
$E_{51}=1$, $n_{0}=1 \,cm^{-3}$ and $c_{0,6}\simeq 1$ 
we obtained the following range of possible values for $\eta_{SN}$:
\begin{eqnarray}
  \eta_{SN}= 0.13 \div 0.007
\end{eqnarray}
In particular, we have adopted an intermediate value 
of $\eta_{SN}=0.03$ which we assume to be the typical 
SN energy transfer efficiency.
\par
The energy injected into the ISM by a typical massive star 
through stellar winds during all her lifetime is estimated to be: 
\begin{eqnarray}
  E_{w}=L_{w} \, t_{MS}
\end{eqnarray}
where $t_{MS}$ is the time the star spends in Main Sequence and
$L_{w}$ is the stellar wind luminosity. 
By using the arguments developed in the appendix, we obtained:
\begin{eqnarray}
  \eta_{w}= 0.3 \, (L_{36})^{2/3} \, n_{0}^{-1/2} \, c_{0,6}^{-5/2}
\end{eqnarray}
if we assume a 20$M_{\odot}$ star as a typical
massive star contributing to the stellar winds
then $L_{36}=L_{w}/(10^{36}\, erg \, s^{-1})=0.03$, 
$t_{MS} = 7.9 \cdot 10^{6} \, yr$ with  
$n_{0}=1 \,cm^{-3}$ and $c_{0,6}=1$, 
we obtain:
\begin{eqnarray}
  E_{w} \simeq 10^{49} \,\,erg \,\,\,\, \, and \,\,\,\,\, \eta_{w}=0.03
\end{eqnarray}
Therefore, in our formulation we adopted:
\begin{eqnarray}
 \epsilon_{SN}=0.03 \cdot 10^{51} \,\, erg
\end{eqnarray}
and
\begin{eqnarray}
   \epsilon_{w}= 0.03 \cdot  10^{49} \,\,erg
\end{eqnarray}
It is worth noting that the assumed value for $E_{w}$ is
in agreement with that calculated by Gibson (1994) for a 
star of initial mass $M=20M_{\odot}$. 

\par
The introduction of dark matter halos with variable amounts and 
concentrations of dark matter is considered when we compute the 
binding energy of interstellar gas, $E_{g}^{b}(t)$:
\begin{eqnarray}
  E_{g}^{b}=W_{L}(t)+W_{LD}(t) 
\end{eqnarray}
The two terms on the right of the equation take in account the gravitational 
interaction between the gas mass $M_{g}(t)$, and the total luminous mass 
of the galaxy $M_{L}(t)$, and between the gas mass and
the dark matter $M_{d}$:
\begin{eqnarray}
  W_{L}(t)=-0.5 \,\, G \,\, {M_{g}(t)M_{L}(t) \over r_{L}} 
\end{eqnarray}
\begin{eqnarray}
  W_{LD}(t)=-G \,\, \tilde{w}_{LD} \, {M_{g}(t)M_{d} \over r_{L}}
\end{eqnarray}
where  \,\,\, $ \tilde{w}_{LD} \simeq {1 \over 2\pi} S [1+1.37S] $.  \\

\noindent
$G$ is the gravitational constant, 
and $S=r_{L}/ r_{d}$ is the ratio between the effective radius of luminous
matter and the effective radius of dark matter. 
These equations are taken from Bertin et al. (1992) and are valid for
$S$ defined in the range $0.10 \div 0.45$. \\
It is worth noting that the original formulation of Bertin et al. (1992) was
thought for massive elliptical galaxies, and that it is not necessarily the 
right one for dwarf irregulars. However, we used such a formulation since 
theoretical formulations for the binding energy of dwarf irregulars are 
not available.

\section{Model results}

Our study of the evolution of BCG and DIG 
starts from
the results obtained by Marconi et al. (1994):
{\em i)} the star formation is assumed to proceed in short and 
intense bursts of activity, which may induce galactic winds;
{\em ii)} metal enriched and normal winds have been considered. In 
particular,
for the metal enriched winds the assumption is made that galactic 
winds carry mostly the nucleosynthesis products of supernovae of type II, 
with the consequence of removing elements such as oxygen but 
not elements such as helium and nitrogen which are 
mainly produced in low and intermediate mass stars, and 
ejected through stellar winds. 
\par 
The novelty is that {\em i)} 
galactic winds are powered both by supernova explosions (SNII 
and SNIa) and stellar winds from massive stars, and that {\em ii)} we 
consider the presence of dark matter halos in these
galaxies.
\par
In order to understand the observed distribution of N/O, C/O 
{\em versus} O/H, and of [O/Fe] {\em versus} [Fe/H], we  
computed different galaxy models by varying some parameters
such as the number of bursts, the duration of each burst, the star 
formation efficiency, the galactic wind efficiency, the dark matter 
mass and distribution, and finally, the IMF exponent. 
\par
We first considered three sets of {\em standard models} characterized 
by 1, 3, 5, 7 and 10 bursts of a duration of 20, 60 and 100 $Myr$, 
whereas the other parameters were fixed by the results presented 
in Marconi et al. (1994). In particular, the IMF is the Salpeter one, 
and the star formation efficiency is $\Gamma=1 \, Gyr^{-1}$.
The galactic wind is differential, namely $w_{i}$ 
is different from zero only for the $\alpha$-elements ejected 
through type II supernova explosions. In particular, for the elements
studied here, $\, w_{i} \,$ is zero for N and He, whereas it is 
$w_{i}=\, 10 $ for O, Ne, Na, Mg and Si, is 
$w_{i}= 0.97 \cdot 10 $ for C, since carbon is partly produced and 
ejected by low and intermediate mass stars through stellar winds, 
and is $w_i=  0.3 \cdot 10$ for 
Fe since only $\simeq 30\%$ of this element is produced by SNe II. 
The ratio between dark and luminous matter is assumed to be 10
and that between the luminous and dark core radius is $S=0.3$
for the three sets of standard models.
Normal galactic winds instead require $\, w_{i}=$ to be the same for all 
the chemical elements. 
From now on we will indicate the models with differential winds by a 
capital $D$,
(e.g. $w_i= 10D Gyr^{-1}$ indicating the value for the $\alpha$-elements).
\par

\begin{figure}
\centerline{\psfig{figure=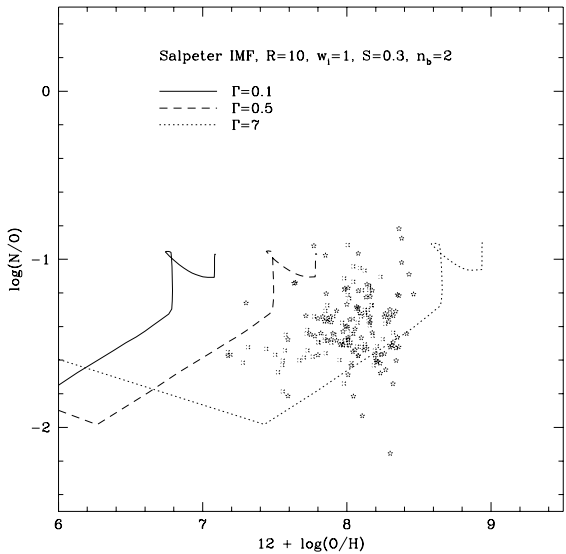,height=8cm}}
\caption[]{Two 0.06 $Gyr$ bursts models: Salpeter IMF, $R=10$, $S=0.3$,
$w_i=1D Gyr^{-1}$. 
We present the observed distribution of $log(N/O)\, vs\, 12 + 
log(O/H)$ and the 
variation range of the star formation efficiency of our models:
$\Gamma=0.1 \div 7 \, Gyr^{-1}$. The best fit however is given by
$\Gamma=0.5 \div 7 \, Gyr^{-1}$.} 
\end{figure}
\begin{figure}
\centerline{\psfig{figure=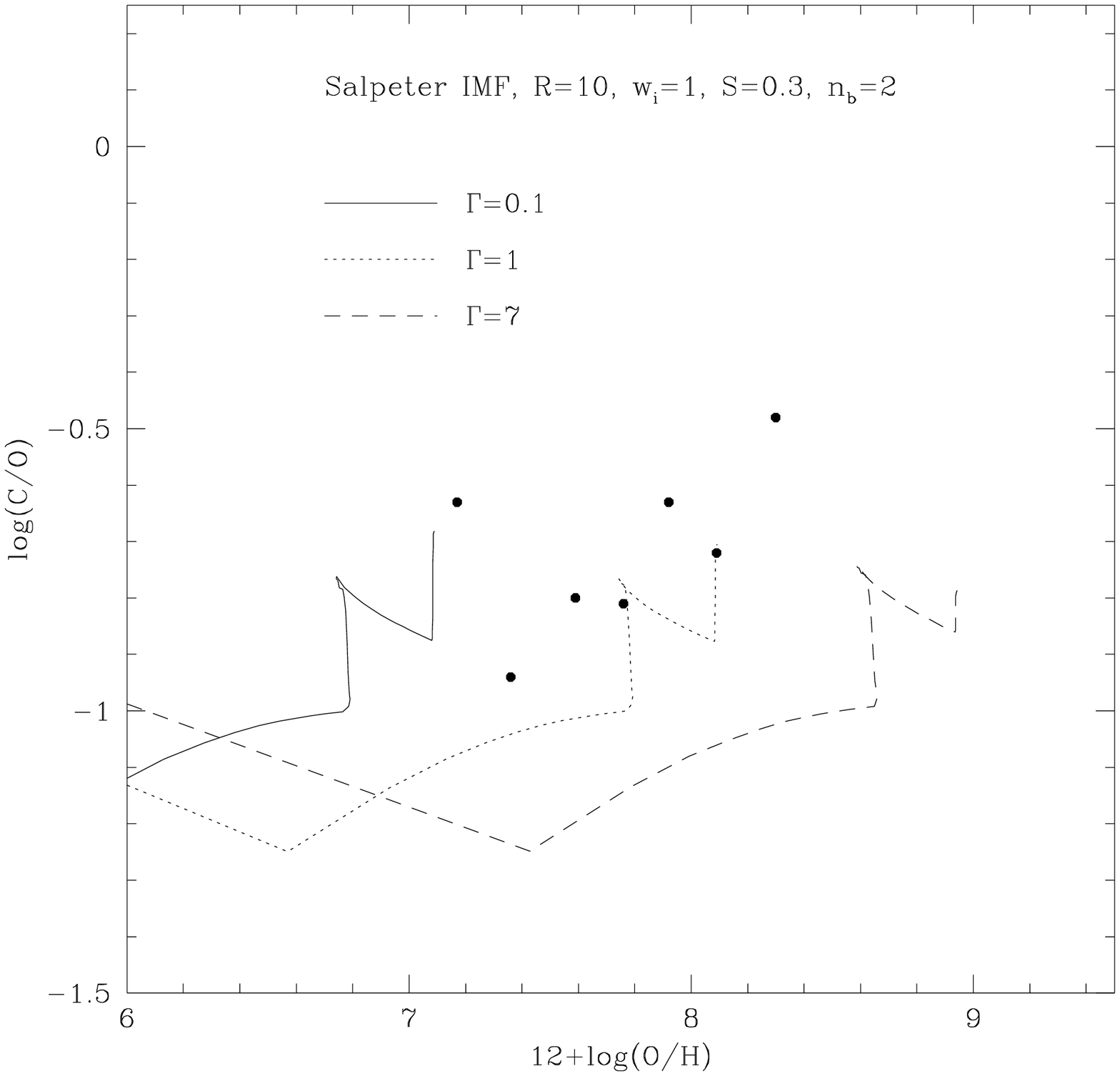,height=8cm}}
\caption[]{Two 0.06 $Gyr$ bursts models: Salpeter IMF, $R=10$, $S=0.3$,
$w_i=1D Gyr^{-1}$. We present the observed 
distribution of $ log(C/O)\, {\em vs} \,
12 + log(O/H) $ and the variation range of the star formation efficiency 
of our models: $\Gamma=0.1 \div 7 \, Gyr^{-1}$. The best fit however is
given by $\Gamma=0.1 \div 1 \, Gyr^{-1}$.} 
\end{figure}
\begin{figure}
\centerline{\psfig{figure=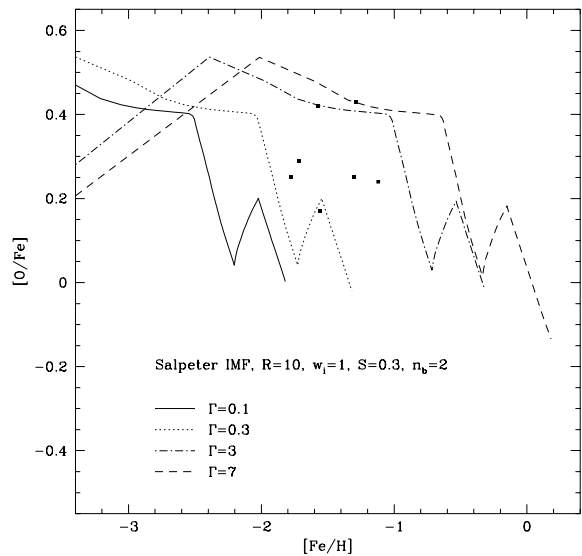,height=8cm}}
\caption[]{Two 0.06 $Gyr$ bursts models: Salpeter IMF, $R=10$, $S=0.3$,
$w_i=1D Gyr^{-1}$. We present the observed distribution of 
$ [O/Fe]\, {\em vs} \,[Fe/H] $ and the variation range of the star 
formation efficiency of our models: $\Gamma=0.1 \div 7 \, Gyr^{-1}$. 
The best fit however is given by $\Gamma=0.3 \div 3 \, Gyr^{-1}$.}  
\end{figure}
\par
Among our standard models, those with burst duration $\Delta t_{b}=20 \,\, 
Myr$ never developed galactic winds, and those with 
$\Delta t_{b}=100 \,\,Myr$ 
predicted the destruction of the galaxy (i.e. when the thermal 
energy of gas is larger than the binding energy of the galaxy)
when a large number of bursts was assumed 
(in particular when $n_{b} \geq 7$); on the contrary models with
$\Delta t_{b}=60 \,\,Myr$ never predicted galaxy destruction 
and developed galactic winds for $n_{b} \geq 5$. 
As a consequence of this, we chose to examine only models
with $\Delta t_{b}=60 \,\,Myr$. 
\par
We started then from our standard models and varied several parameters 
such as the star formation efficiency, the wind efficiency,
the amount and distribution of dark matter.
We found that in order to reproduce the observed spread of $log(N/O)$ and 
$log(C/O) \,\, versus \,\, 12 + log(O/H)$ and $[O/Fe]\,\, versus \,\, 
[Fe/H]$, the most important parameter is the star formation efficiency.
We computed two series of models characterized 
by two different IMF: the Salpeter (1955) and the Scalo (1986) function. 
To reproduce the observed spread in all three diagrams, we needed a
star formation efficiency ($\Gamma$) between 0.1 and 7 $Gyr^{-1}$ 
when using the Salpeter IMF, and between 1 and  10 $Gyr^{-1} $ when using the 
Scalo IMF. The best value for the wind parameter was $w_i=1D Gyr^{-1}$.
\par
Consider first the Salpeter IMF case: we computed models with 1, 2, 3, 5 
and 7 bursts, and tested the importance of the presence of the dark matter. 
Greater is the amount of dark matter, deeper is the galaxy potential 
and higher the gas binding energy. Therefore, the thermal energy required to 
develop the galactic wind has also to be higher.
We found that if the amount of dark matter is the same as that of the
luminous matter ($R=M_{d}/M_{L}=1$) any model predicts galaxy destruction
during the first starburst
if the star formation efficiency is high ($\Gamma \geq 5$).
This upper limit of $\Gamma$ depends of course on the number of bursts:
models with many bursts, are characterized by a great consume of gas
through star formation activity, so they must have lower star formation
rates to keep bound, than models with just one or two bursts.
We previously said that we need $\Gamma =0.1 \div 7 \,\,Gyr^{-1}$ to 
explain the observed spread of the abundances ratios. 
So, let's consider the presence of a greater amount of dark matter: 
ten times the luminous matter ($R=10$). In this case
none of the models characterized by 1 or 2 bursts and 
$\Gamma =0.1 \div 7 \,\, Gyr^{-1}$ predicts galaxy destruction
during starbursts, whereas 
those with 3 or more bursts sometimes do. 
Considering an even greater amount of dark matter, fifty times 
the luminous matter ($R=50$), none of the models predicts galaxy destruction 
for any value of the star formation efficiency 
$\Gamma=0.1 \div 7 \,\, Gyr^{-1}$.
\par
In table \ref{t1} we summarize the range of variation of the star
formation efficiency and the occurrence of the galactic wind for 
different values of the number of bursts and different amounts of
dark matter for stable models (i.e. the galaxy never blows up). 
In column 1 is presented the number of bursts
($n_{b}=$ 1, 2, 3, 5, 7), in column 2 the range of variation of the 
star formation efficiency
($\Gamma$) and in column 3 we indicate if models develop the galactic wind
or not. This same scheme is reproduced three times, for $R=1,\, 10$ and 50.

\begin{table}
\caption{Salpeter IMF models characterized by  $S=0.3$, $w_i=1D Gyr^{-1}$: 
we indicate the range of variation of the star formation efficiency 
($\Gamma$, expressed in $Gyr^{-1}$) for different values of $R$ and 
$n_{b}$, and the development of the galactic wind ($GW$).} 
\vspace{0.5cm}
\begin{tabular}{|c||c|c||c|c||c|c|}  \hline
  & \multicolumn{2}{c||}{$R=1$} & \multicolumn{2}{c||}{$R=10$} & 
   \multicolumn{2}{c|}{$R=50$}   \\ \hline \hline

 $n_{b}$ & $\Gamma$ & $GW$ & $\Gamma$ & $GW$ & $\Gamma$ & $GW$   
         \\ \hline \hline

 1 & $0.1 \div 5$ & yes  & $0.1 \div 7$ & no & $0.1 \div 7$ & no 
    \\ \hline
 2 & $0.1 \div 3$ & yes & $0.1 \div 7$ & yes & $0.1 \div 7$ &  no
    \\ \hline
 3 & $0.1 \div 1$ & yes & $0.1 \div 5$ & yes & $0.1 \div 7$ & no 
   \\ \hline
 5 & $0.1 \div 1$ & yes  & $0.1 \div 3$ & yes & $0.1 \div 7$ & yes
   \\  \hline
 7 & $\leq 0.1 $ & yes & $0.1 \div 1$ & yes  & $0.1 \div 7$ & yes  
   \\ \hline 
\end{tabular}
\label{t1}
\end{table}
\par
Therefore, one of the main results of our study is the fact that the
presence of dark matter halos around dwarf irregulars and blue 
compact galaxies is required in order to avoid total destruction due 
to the energy injected by supernova explosions and stellar winds during 
starbursts. 
Moreover, the distribution of dark matter in our formulation, is 
described by the parameter $S=r_{L}/r_{d}$, and we find that $S$ 
can vary between 0.1 and 0.4. 
\par
Figure 1 shows the range of variation of the star formation 
efficiency for standard models characterized by two 0.06 $\, Gyr$ bursts
(occurring at $t=1,\,$ and $5 \,\,Gyr$), $R=10$, $S=0.3$, 
$w_i=1D Gyr^{-1}$ and the Salpeter IMF.
We notice the peculiar sawtooth behaviour, indicating 
the alternation of the active star formation periods (during the bursts) 
and the quiescent periods (during the interbursts).
During the starbursts, the star formation rate is high, and the SNe of 
type II dominate in the chemical enrichement of O, Fe, N and C, while 
during the interbursts, when the star formation is not acting, only the 
elements like C, N, or Fe, are produced by low and intermediate mass stars. 
In this case the abundance of these elements increases relative to the
oxygen abundance, and this is exactly what we observe in all the
figures presented here.
The observed spread of N/O abundance ratio, reported in figure 1, 
is quite well reproduced if $\Gamma=0.5 \div 7 \,Gyr^{-1}$. 
\par
Figures 2 and 3 show the C/O and the [O/Fe] abundance ratios
relative to the same models, respectively. In these cases we can notice how 
the observed 
spreads are quite well reproduced if $\Gamma=0.1 \div 1 \,\,Gyr^{-1}$ for
C/O and if $\Gamma=0.3 \div 3 \,\,Gyr^{-1}$ for [O/Fe]. 
These differences are probably due to the few data available in
literature for C and Fe
abundances and to 
uncertainties present in the nucleosynthesis calculations. 
\par
On the other hand, figures 4, 5 and 6 show the
same models but with the Scalo IMF, reproducing the N/O, C/O and [O/Fe] 
abundance ratios. In this case the best results are obtained with 
$\Gamma=1 \div 10 \,\,Gyr^{-1}$. 
However, while the [Fe/O] observed 
spread is quite well reproduced (figure 6), the C/O and in 
particular the N/O abundance ratio are worsely reproduced than by the
Salpeter IMF models (figures 2  and 1). 
\par
On the basis of our study we can conclude that models with the Salpeter IMF 
are favoured relative to those adopting the Scalo one, 
since these latter do not explain all the spread present in the N/O vs O/H 
diagram. The models with the Salpeter IMF
require a star formation 
efficiency $\Gamma=0.1 \div 7 \,\, Gyr^{-1}$ and a dark to luminous 
matter ratio $R=1 \div 50$.
\par
In our analysis of the best model we have also 
considered the relation between metallicity ($Z$)
and the gas mass fraction ($\mu=M_{gas}/M_{tot}$ with 
$M_{tot}=M_{L}+M_{d}$). Matteucci and Chiosi (1983) discussed the 
problem of reproducing the observed spread existing in the $Z-\mu$
diagram and suggested that different wind rates, different infall rates and 
different IMF from galaxy to galaxy
could equally well explain the spread. 
However, they did not consider the role played by the dark matter.
On the other hand, Kumai and Tosa (1992) 
suggested that the observed spread could be explained
with the presence of variable amounts of dark matter:
they proposed that the dark matter fraction should vary from galaxy to galaxy
as  $f_{D}=M_{d}/M_{tot}=0.40 \div 0.95$.
Our results agree with Kumai and Tosa (1992) and, 
as one can see in figure 7,
where it is shown that the dark matter can vary between 1 and 50 
times the luminous
matter ($R=1 \div 50$) in order to reproduce the spread.
This means that the parameter $f_{D}$ should vary in the range
$f_{D}=0.50 \div 0.98$.
\begin{figure}
\centerline{\psfig{figure=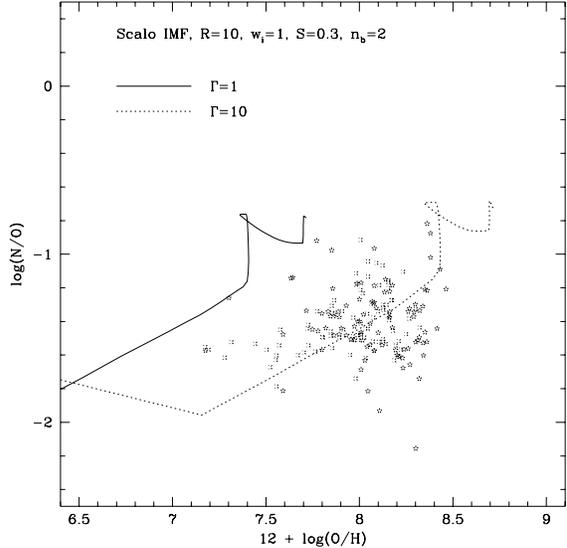,height=8cm}}
\caption[]{Two 0.06 $Gyr$ bursts models: Scalo IMF, $R=10$, $S=0.3$,
$w_i=1D Gyr^{-1}$. We present 
the observed distribution of $ log(N/O)\, {\em vs} \,
12 + log(O/H) $ and the variation range of the star formation efficiency 
of our models: $\Gamma=1 \div 10 \, Gyr^{-1}$.} 
\end{figure}
\begin{figure}
\centerline{\psfig{figure=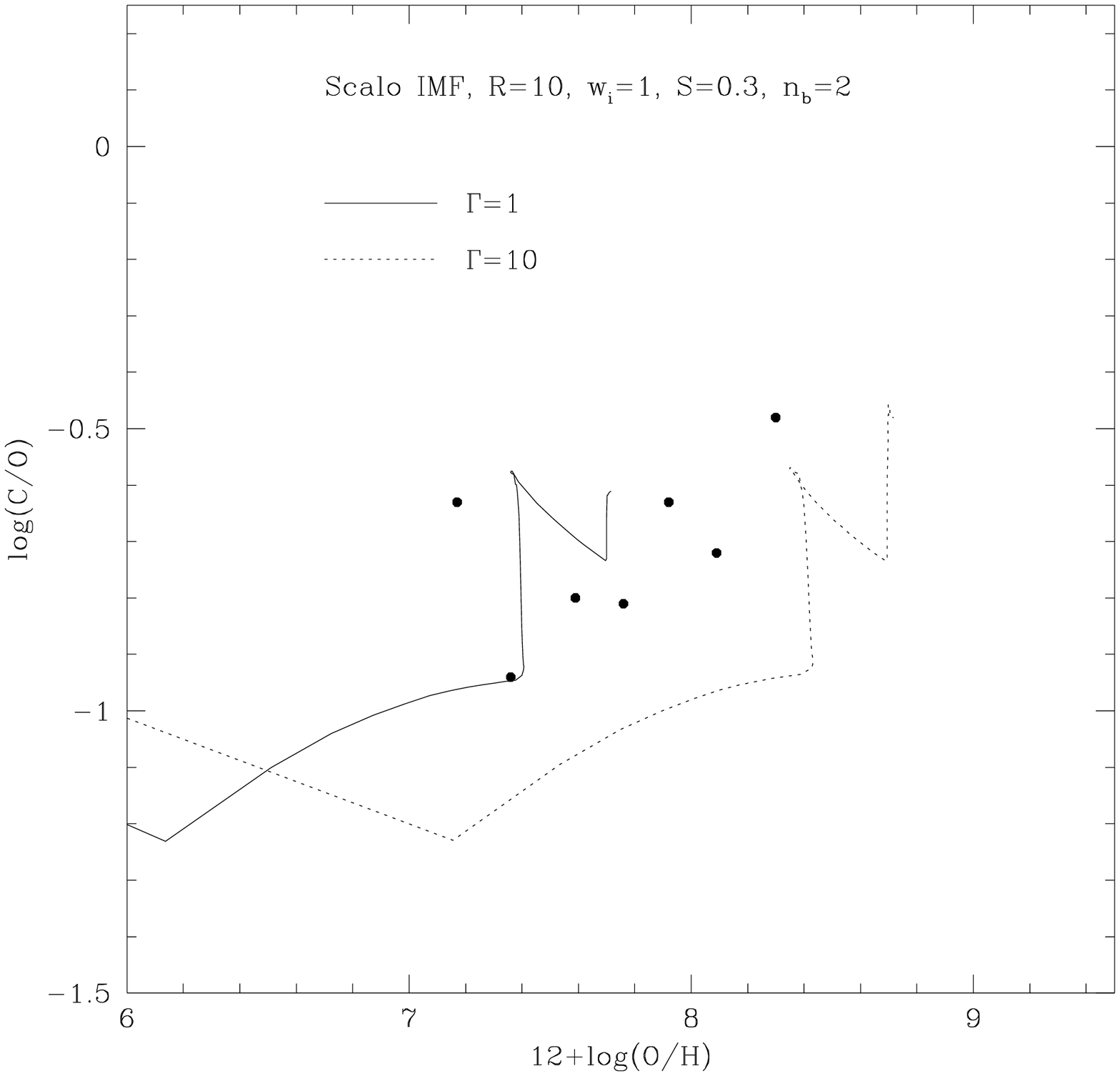,height=8cm}}
\caption[]{Two 0.06 $Gyr$ bursts models: Scalo IMF, $R=10$, $S=0.3$,
$w_i=1D Gyr^{-1}$. We present the observed distribution of $ log(C/O)\, 
{\em vs} \,
12 + log(O/H) $ and the variation range of the star formation efficiency 
of our models: $\Gamma=1 \div 10 \, Gyr^{-1}$.} 
\end{figure}
\begin{figure}
\centerline{\psfig{figure=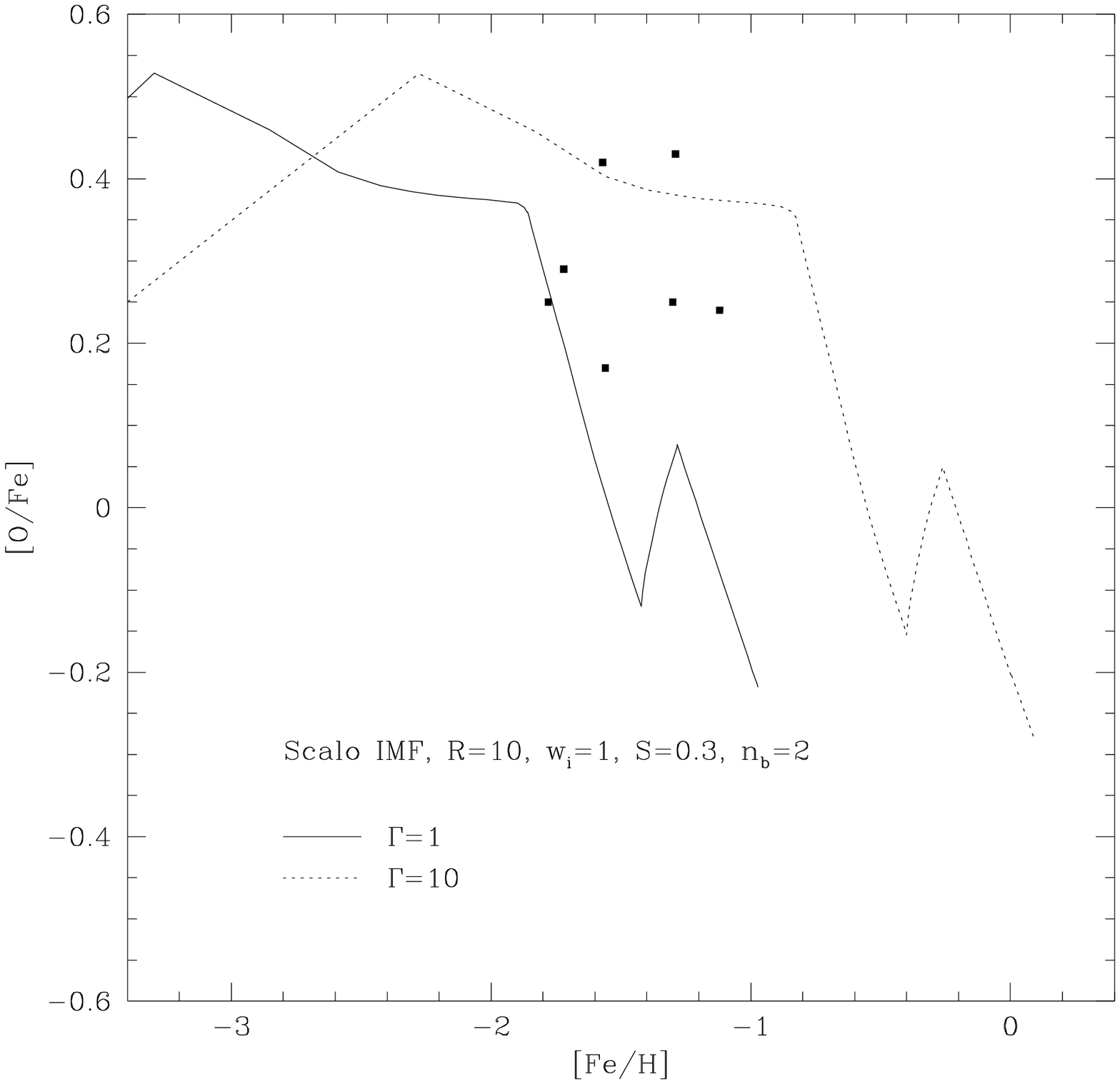,height=8cm}}
\caption[]{Two 0.06 $Gyr$ bursts models: Scalo IMF, $R=10$, $S=0.3$,
$w_i=1D Gyr^{-1}$. We present the observed distribution of $ [O/Fe]\, {\em vs} \,
[Fe/H] $ and the variation range of the star formation efficiency 
of our models: $\Gamma=1 \div 10 \, Gyr^{-1}$.} 
\end{figure}
\begin{figure}
\centerline{\psfig{figure=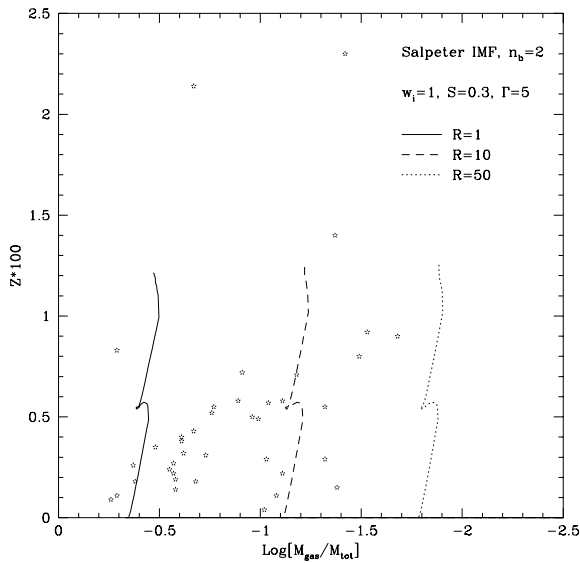,height=8cm}}
\caption[]{Two 0.06 $Gyr$ bursts models: Salpeter IMF, 
$\Gamma=5 \,\, Gyr^{-1}$$S=0.3$, $w_i=1D Gyr^{-1}$. We present the observed 
distribution of $Z$ {\em versus} $Log \mu$ ($\mu=M_{gas}/M_{tot}$). 
The models are
characterized by a different amount of dark matter: $R=1, 10, 50$.}
\end{figure}

\par
The second main result of our study concerns the energetics of 
interstellar gas: the ISM receives energy from both supernova 
explosions and stellar winds from massive stars. 
In figure 8 we can notice the interstellar gas thermal energy
relative to its binding energy and the total galaxy 
binding energy. When the thermal energy equates the gas binding energy, the
galaxy develops the galactic wind.
When a supernova explodes it injects almost
$10^{51}\,\, erg$ into the ISM, but just some percents (see paragraph 3.) 
of this energy are transformed into thermal energy of the gas.
The question about stellar winds from massive stars is still under debate. 
As already discussed,
we considered here 
a typical massive star ($ M\simeq 20M_{\odot}$) and assumed that it 
may inject into the ISM something like $0.03 \cdot 10^{49} \,\,erg$ through 
stellar winds during all of its life, 
and our results in this case suggest
that the total thermal energy due to stellar 
winds from massive stars, is negligible if compared to the component 
due to supernovae of type II and Ia. 
Supernovae of type Ia also do not contribute
significantly, as shown in  
Figure 9 where the different contributions to the 
thermal energy due to the SNeII, the SNeIa and the stellar winds are presented.
This would mean that both stellar 
winds and SNeIa play a negligible role in the evolution of these systems.
\par

\begin{figure}
\centerline{\psfig{figure=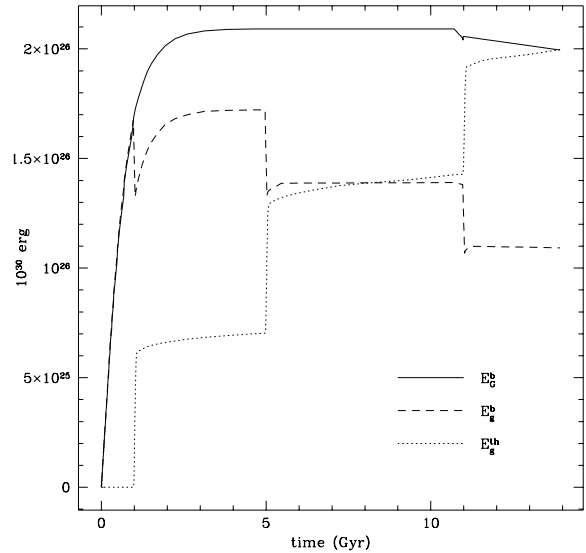,height=8cm}}
\caption[]{Three 0.06  $Gyr$ bursts model, characterized by: Salpeter 
IMF, $\Gamma=5$, $R=10$, $S=0.3$, $w_i=1D Gyr^{-1}$. The bursts occur at 
$t=1, 5$ and 11$Gyr$.
We present the galaxy binding energy ($BE_{G}$), the interstellar gas
binding energy ($E_{g}^{b}$) and the interstellar gas thermal energy
($E_{th}$). The time of occurence of the galactic wind correspond to the
time at which $E_{th} = E_{g}^{b}$.}
\end{figure}
\begin{figure}
\centerline{\psfig{figure=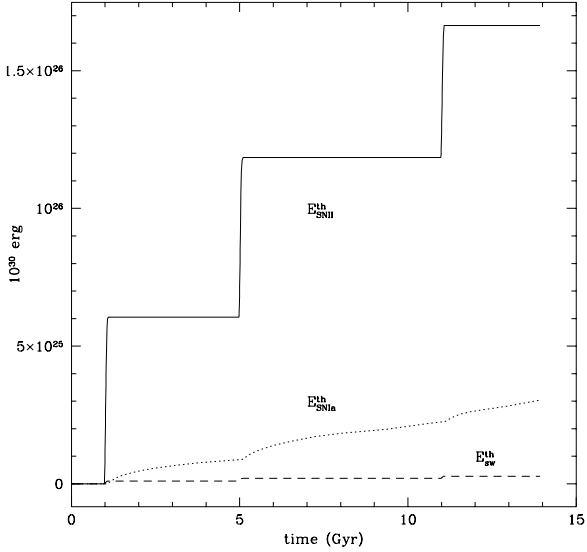,height=8cm}}
\caption[]{Three 0.06  $Gyr$ bursts model, characterized by: Salpeter 
IMF, $\Gamma=5$, $R=10$, $S=0.3$, $w_i=1D Gyr^{-1}$. 
The bursts occur at $t=1, 5$ and 11
$Gyr$. We present the SNeIa and the stellar winds contributions 
to the total thermal energy relative to the SNeII contribution:
note that the stellar winds contribution is smaller than the one
due to the SNeIa.}
\end{figure}

\par
In agreement with the Marconi et al. (1994) results, our results
also favour differential
galactic winds, but we find that the wind efficiency parameter $W$
has to be lower ($W \leq 1 \,\,Gyr^{-1}$) than in Marconi et al. (1994).
\par
Finally in figure 10 is reported the characteristic 
behaviour of type II supernova rates:
each peak corresponds to a burst of star formation. 
In figure 11 instead typical type Ia 
supernova rates are presented and we can notice how type Ia supernovae 
explode also during the interbursts periods. 
Considering Salpeter models characterized by $R=50$, $S=0.3$, and $w_i=1D 
Gyr^{-1}$, 
we find that the present
value of SNeIa rate varies between $0.32 \cdot 10^{-9}$ and $0.67 \cdot
10^{-4} \,\, yr^{-1}$, depending on the values of both the star 
formation efficiency and the number of bursts.
In particular, in table \ref{t2} we indicate the range of variation of the 
present value of type Ia SNe rate ($R_{SN Ia}$) for different values of
the number of bursts ($n_{b}=1, 2, 3, 5, 7$) and 
$\Gamma=0.1 \div 7 \,\, Gyr^{-1}$.
\begin{figure}
\centerline{\psfig{figure=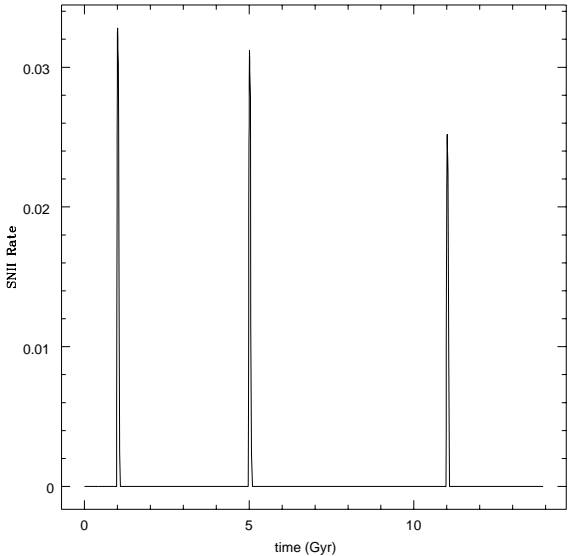,height=8cm}}
\caption[]{Three 0.06  $Gyr$ bursts model, characterized by: Salpeter 
IMF, $\Gamma=5$, $R=10$, $S=0.3$, $w_i=1D Gyr^{-1}$. 
The bursts occur at $t=1, 5$ and 11 $Gyr$. We present the rate of type 
II supernova explosion as a function of time. The units of the SN rate 
are $SNe yr^{-1}$}
\end{figure}
\begin{figure}
\centerline{\psfig{figure=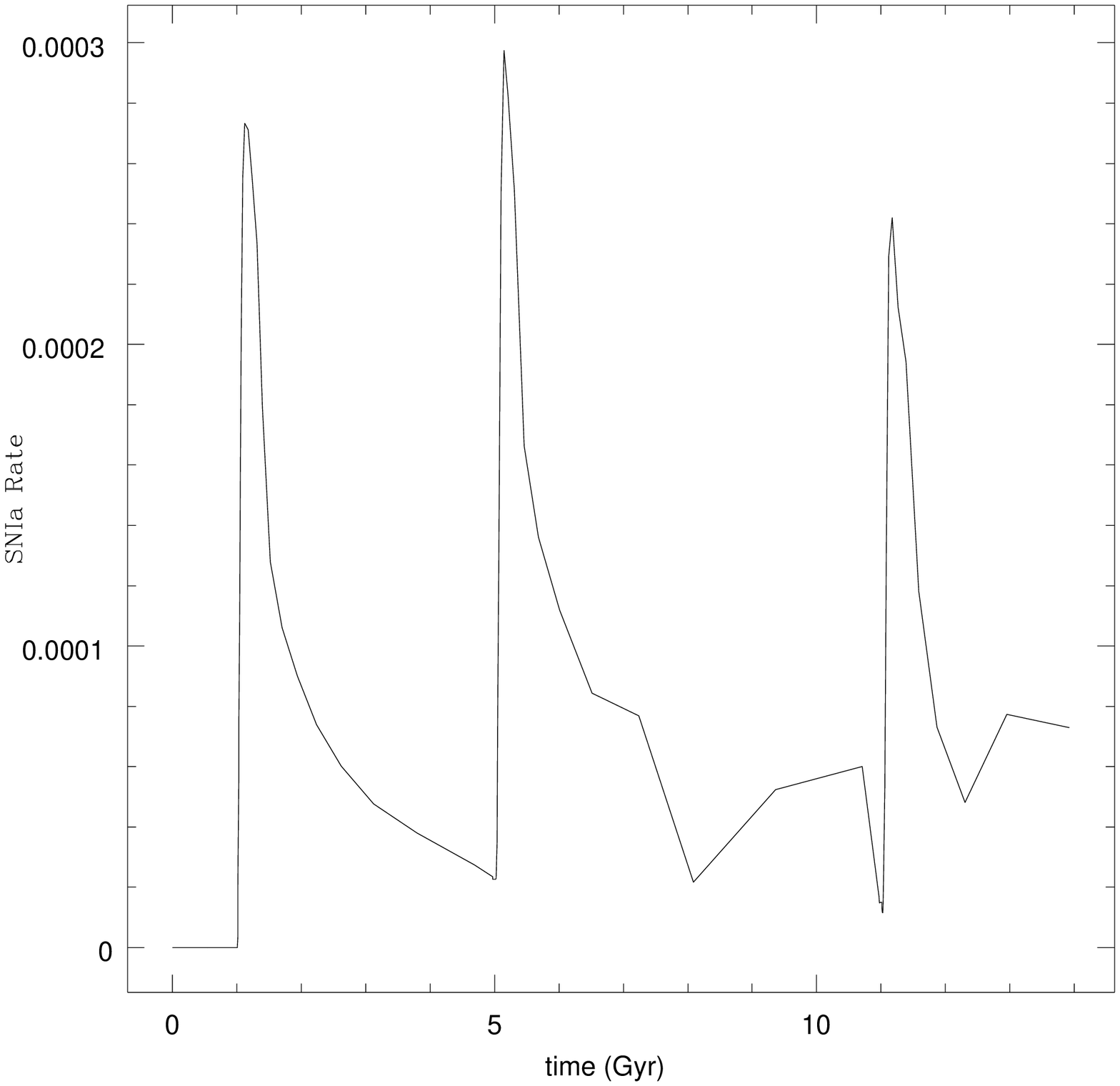,height=8cm}}
\caption[]{Three 0.06  $Gyr$ bursts model, characterized by: Salpeter 
IMF, $\Gamma=5$, $R=10$, $S=0.3$, $w_i=1D Gyr^{-1}$. The bursts occur at 
$t=1, 5$ and 11 $Gyr$. We present the rate of type Ia supernova explosion 
as a function of time. The units of the SN rate are $SNe yr^{-1}$}
\end{figure}
\begin{table}
\caption{Range of variation of type Ia supernova rates ($R_{SN Ia}$) 
for different values of the number of bursts ($n_{b}$). The
star formation efficiency is $\Gamma=0.1 \div 7 \, Gyr^{-1}$.}
\vspace{0.5cm}
\begin{tabular}{|l||c|}  \hline
   $n_{b}$ & $R_{SN Ia} \,\, (yr^{-1})$ \\ \hline \hline

 1 & $0.32 \cdot 10^{-9} \div 0.80 \cdot 10^{-5}$  \\ \hline
 2 & $0.37 \cdot 10^{-6} \div 0.31 \cdot 10^{-5}$  \\ \hline
 3 & $0.24 \cdot 10^{-5} \div 0.82 \cdot 10^{-4}$  \\ \hline
 5 & $0.25 \cdot 10^{-5} \div 0.68 \cdot 10^{-4}$  \\ \hline
 7 & $0.30 \cdot 10^{-5} \div 0.67 \cdot 10^{-4}$  \\ \hline
\end{tabular}
\label{t2}
\end{table}

\section{Summary}

In this paper we attempted to model the chemical evolution of 
dwarf irregulars and blue compact galaxies.
In particular, we considered the presence of dark matter halos and of 
the different contributions to the interstellar gas energy due to 
supernovae (II and Ia) and stellar winds from massive stars.
The comparison with the available
data on the abundances and abundance ratios of 
elements such as He, N, C, O, and Fe allowed us to conclude that 
DIG and BCG 
must contain  a substantial amount of dark matter in 
order to be gravitationally bound against the intense starbursts, 
that galactic winds powered by supernovae and stellar winds from massive
stars are preferably enriched (differential), and that the 
energy contribution from type Ia supernovae to the total thermal energy of the 
gas is negligible relative to the type II supernova 
component. We found also that the stellar wind contribution is negligible 
relative to that of the type II supernovae. 

\par
In summary, our models suggest:
\begin {itemize}
\item
a number of bursts $n_{b}=1 \div 10$;
\item
a star formation efficiency $\Gamma=0.1 \div 7 \, Gyr^{-1}$;
\item
differential galactic winds mostly powered by supernovae of type II, with a 
wind efficiency parameter $W \leq 1 \, Gyr^{-1}$;
\item
a ratio between dark matter and luminous matter $R=1\div 50$;
\item
a dark matter distribution given by the ratio between luminous effective radius
and dark effective radius $S= 0.1 \div 0.4$;
\item
Salpeter IMF is favoured relative to Scalo IMF: in fact the observed 
abundance ratios are better reproduced by Salpeter models than by 
Scalo models, in particular the spread of N/O abundance ratio.
\end{itemize}

\vspace{2cm}
\def\acknowledgements{\noindent{\it Acknowledgements.\ }}
\begin{acknowledgements}
We thank Fabrizio Brighenti and C. Chiappini for many useful discussions.
\end{acknowledgements}
\bigskip\bigskip
{\bf Appendix} \par
\bigskip
In order to calculate the energy gained by the ISM after an isolated SN
explosion we consider the expansion of the SNR through an uniform medium of
number density $n_0$ and isothermal sound speed $c_0=10^6c_{0,6}$ cm s$^{-1}$.
Initially the temperature of the remnant is very high and the
energy radiated away is negligible as compared to the explosion energy $E_0$;
this phase is called `adiabatic' and is described by the Sedov solution,
$\rs \propto t^{2/5}$.
As the remnant expands its temperature decreases and eventually the radiative
losses from the swept-up dense shell become relevant at the cooling time 
given by (cf. Cioffi \& Shull 1991):

$$\tc=1.49\times 10^4 {E_{51}^{3/14}\over n_0^{4/7}\zeta^{5/14}}\;{\rm yr}$$
\noindent
where $E_0=10^{51}E_{51}$ erg and $\zeta$ is the metal abundance relative to
the solar abundance.

For $t>\tc$
most of the mass and the kinetic energy of the SNR is contained by the
thin, cold, massive expanding shell, while the  interior cavity is filled with
hot dilute gas that contains most of the thermal energy. At this stage
the hot bubble has a very long cooling time and cools adiabatically as
it expands pushing the outer shell. We thus assume that its thermal energy
varies as $E_{\rm th}\propto \rs^{-2}$ and neglect a correction $\propto
t^{-4/7}$ due to the effect of cooling (Cioffi, Mckee \& Bertschinger, 1988).
This is compatible with the other simplifying assumptions made in our
evaluation of the SN efficiency.
The remnant therefore expands as $\rs \propto t^{2/7}$ in this 
`snow plow' phase.  At a later time the pressure inside
the cavity becomes equal to the external pressure and the remnant stalls.
We define the stalling radius as the radius at which
the shell velocity is equal to $c_0$.
As long as the SNR expands supersonically, the ISM outside the remnant
is not aware of its presence and the fraction of the SN explosion energy which
is not cooled off remains locked inside the cavity. At merger, the
shell dissipates its remaining kinetic energy and its thermal energy
(we assume that the shell temperature does not become lower than the
temperature of the unperturbed ISM, so that its thermal energy is three times
the kinetic energy at the stalling radius) which add to the thermal content
of the ISM. The efficiency is thus given by the ratio of this shell
energy to the explosion energy. With the assumptions above it is found:

$$\eta_{\rm SN}=0.12E_{51}^{2/35}n_0^{-4/35}\zeta^{-8/35}c_{0,6}^{4/5}$$
\noindent
It may happen that the remnant pressure becomes lower than the external
pressure before the merger. In this case the remnant enters in the
`momentum conserving' phase, $\rs \propto t^{1/4}$.
Spitzer (1968) has shown that, in the
limiting case in which the `snow plow' phase is entirely skipped and
the `momentum conserving' phase starts just after $\tc$, the efficiency
is given by ${c_0 \over 5v_{\rm c}}$, where $v_{\rm c}$ is the remnant
velocity at $\tc$. In our notations we get:

$$\eta_{\rm SN}=0.005E_{51}^{-1/14}n_0^{-1/7}\zeta^{-3/14}c_{0,6}$$
\noindent
Thus, the `true' efficiency will be intermediate between the two above.
Andersen \& Burkert (submitted), in describing the evolution of the ISM
of the dwarf galaxies, assume $\epsilon_{\rm SN}=0.08$ independent of the
environment, but find that the exact value is not of importance. 

Before to explode, stars earlier than B0 ($M>20 \msun$)
suffer considerable mass loss ($\dot M\sim 10^{-5}-10^{-6}$ $\msun$ yr$^{-1}$)
through supersonic winds $(v_{\rm w}\sim 10^5-10^6$ km s$^{-1}$), carving
large, hot bubbles (Weaver et al., 1977) and injecting large amounts of
energy into the ISM. In fact, as the wind impinges on the ISM, a reflected 
shock forms which thermalizes the wind itself. This rarefied, hot,
pressurizzed gas fills the expanding cavity which is surrounded by a thin,
cold, dense shell formed by the swept up gas of the ISM. As the expansion
procedes the radiative losses increase and the inner pressure
decreases untill eventually the bubble stalls.
As for the SNR case, we calculate the kinetic energy of
the dense shell pushed by the expanding bubble when it becomes sonic. 
To obtain the wind efficiency $\epsilon_{\rm w}$ the
shell energy is then divided by the total mechanical energy of the wind
$t_{\rm MS}L_{\rm w}$, where $t_{\rm MS}$ is the time spent by the star on
the main sequence, and $L_{\rm w}=10^{36}L_{36}$ erg s$^{-1}$ is the wind
mechanical luminosity. From the classical `energy conserving' solution of
Weaver et al., $R\propto t^{3/5}$, we get:

$$\eta_{\rm w}=0.3L_{36}^{2/3}n_0^{-1/2}c_{0,6}^{-5/2}.$$

\noindent
To obtain the above relation we assumed $T_{\rm MS}=4.4L_{36}^{-1/6}$
(McKee et al. 1984). As in the case of SNRs, the above efficiency represents
an upper limit; in fact, `momentum conserving' bubbles ($R\propto t^{1/2}$)
may be realized if
a substantial enhancement of radiative losses occurs for some reason such as 
cloudy medium (McKee et al. 1984) or non steady cooling (e.g. D'Ercole 1992).
In this case the hot gas quickly cools and the shell is pushed by the ram
pressure of the wind that impinges directly on it. The efficiency then 
becomes:

$$\eta_{\rm w}=4.7\times 10^{-4}L_{36}^{2/3}n_0^{-1/2}V_{2000}^{-3/2}c_{0,6}^{-1},$$

\noindent
where $V_{2000}$ is the wind velocity in units of $2\times 10^8$ cm s$^{-1}$

We finally point out that in principle the SN efficiency worked out
above could not be valid for exploding stars more massive than 20 $\msun$
because the remnant interacts with the bubble rather than with the
uniform ISM. If however we consider 20 $\msun$ as representative of
massive stars ($L_{36}=0.03)$, then the maximum bubble radius is rather
small ($\sim 10$ pc), shorter of the radius where the SNR becomes
radiative ($\sim 15$ pc) and much shorter than the radius at which it stalls
($\sim 60$ pc). If, moreover, heat conduction is active, the bubble tends to
be replenished during the red supergiant phase of the star (D'Ercole 1992).
We thus consider the above estimate of $\epsilon_{\rm SN}$ rather accurate
for our purposes.

\bigskip\bigskip

{\bf References}

\bigskip

\par\noindent
 Angeletti, Giannone 1990, Astron.Astroph., 234, 53 
\par\noindent
Arimoto N., Yoshii Y. 1987, Astron.Astroph., 164, 260 
\par\noindent
Bertin G. et al. 1992, Ap.J., 384, 423 
\par\noindent
 Bl\"ocker T., Sch\"onberner D., 1991, Astr.Astroph., 244, L43
\par\noindent
Bressan A. et al. 1994, Ap.J. Suppl., 94, 63 
\par\noindent
Cioffi D.F., McKee C.F., Bertschinger E., 1988, ApJ, 334, 252 
 \par\noindent
Cioffi D.F., Shull J.M., 1991, Ap.J. 367, 96 
 \par\noindent
Ciotti L., D'Ercole A., 1989, Astr.Astroph., 215, 347
\par\noindent
 Cox D.P., 1989, in {\em Structure and Dynamics of the Interstellar medium}, 
IAU Coll No 120, eds. Tenorio-Tagle G., Moles M., Melnick J., p. 500
\par\noindent
D'Ercole A., 1992, M.N.R.A.S., 255, 272 
\par\noindent
Dufour R.J., Hester J.J., 1990, Ap.J., 350, 149
\par\noindent
Dufour R.J., Garnett D.R., Shields G.A., 1988, Ap.J., 332, 752
\par\noindent
Ferrini F., Poggianti 1993, Ap.J., 410, 44 
\par\noindent
 Gerola H., Seiden P.E., Schulman L.S., 1980, Ap.J., 242, 517
\par\noindent
 Garnett D.R 1990, Ap.J., 363, 142 
\par\noindent
Garnett D.R et al. 1995a, Ap.J., 443, 64 
\par\noindent
Garnett D.R et al. 1995b, Ap.J., 449, L77 
\par\noindent
Gibson B.K.1994, M.N.R.A.S., 271, L35
\par\noindent
 Hensler G., Theis C., Burkert A., 1992, in {\em The feedback of chemical 
evolution on the stellar content of galaxies}, 3rd DAEC Meeting, eds: 
Alloin D., Stasiqska, p229
\par\noindent
 Huchra J.P., 1977, Ap.J.Suppl., 35, 161
\par\noindent
 Klein U., Wielebinski R., Thuan T.X., 1984, Astr.Ap., 141, 241
\par\noindent
Kumai X. and Tosa X. 1992, Astron.Astroph., 257, 511 
\par\noindent
 Kunth D., Sargent W.L.W., 1986, Ap.J., 300, 496
\par\noindent 
 Kunth D., Lequeux J., Sargent W.L.W., Viallefond F., 1994, Astr.Ap., 282, 709
\par\noindent
Kunth D., Matteucci F. and Marconi, G., 1995, Astron.Astrophys., 297, 634
\par\noindent
 Lequeux J., Viallefond F., 1980, Astr.Ap., 91, 269
\par\noindent
 Lequeux J., Peimbert M., Rayo J.F., Serrano A., Torres-Peimbert S., 1979,
              Astr.Ap., 80, 155
\par\noindent
 Lequeux J., Maucherat-Joubert M., Deharveng J.M., Kunth D., 1981, Astr.Ap.,
 103, 305
\par\noindent
Maeder A., Meynet, G., 1989, Astr.Ap., 210, 155
\par\noindent
Maeder A., 1992, Astr.Ap., 264, 105
\par\noindent
 Marconi G., Matteucci F., Tosi M., 1994, M.N.R.A.S., 270, 35
\par\noindent
 Mas-Hesse J.M., Kunth D., 1991, A.A.S.S., 88, 399 
\par\noindent
Matteucci F., Chiosi C. 1983, Astron.Astroph., 123, 121 
\par\noindent
Matteucci F, Tornambe' A. 1987, Astron.Astroph., 185, 51 
\par\noindent
Matteucci F., 1986, M.N.R.A.S., 221, 911
\par\noindent
Matteucci F., Padovani P. 1993, Ap. J., 419, 485
\par\noindent 
 Matteucci F., Tosi M., 1985, M.N.R.A.S., 217, 391
\par\noindent 
 Morton D.C., 1991, Ap.J.S., 77, 119
\par\noindent
Nomoto K., Thielemann F.K., Yokoi K., 1984, Ap.J., 286, 644
\par\noindent
 Pagel B.E.J., Simonson E.A., Terlevich R.J., Edmunds M.G., 1992, M.N.R.A.S.,
 255, 325
\par\noindent 
 Pantelaki I., Clayton D.D., 1987, in Starbursts and galaxy evolution. 
 Thuan T.X., Montmerle T., Tran Thanh Van J. (eds). Editions Frontieres,
 Gif-sur Yvette, p.145
\par\noindent
 Pilyugin L.S., 1992, Astr.Astroph., 260, 58
\par\noindent
Pilyugin L.S., 1993, Astron.Astroph., 277, 42 
\par\noindent
Reimers D., 1975, Mem.R.Sci. Liege 6 eme Ser., 8, 369
\par\noindent
Renzini A., Voli M., 1981, Astr.Astroph., 94, 175
\par\noindent
Roy J.R., Kunth D., 1995, Astron. Astrophys., 294, 432
\par\noindent
 Searle L., Sargent W.L.W., 1972, Ap.J., 173, 25
\par\noindent
 Searle L., Sargent W.L.W., Bagnuolo W.G., 1973, Ap.J., 179, 427
\par\noindent
Skillmann E.D., 1996, preprint
\par\noindent
 Skillman E.D., Kennicutt R.C., 1993, Ap.J., 411, 655
\par\noindent
 Spitzer L. Jr., 1968, {\it Diffuse Matter in Space}, Wiley, New York
\par\noindent
Thielmann et al. 1993, from {\em Origin and evolution of elements} ed. 
N. Prantzos et al., Cambridge Univ. press, p.297 
\par\noindent
Tosi M., Greggio L., Focardi G., Marconi G., 1992, in IAU Symp. {\em 
Stellar populations of galaxies} ed. B. Barbuy e A. Renzini, Kluwer, p. 207 
\par\noindent
 Viallefond F., Lequeux J., Comte G., 1987, in Starburst and galaxy evolution.
Thuan T.X., Montmerle T., Tran Thanh Van J. (eds). Editions Frontieres,
 Gif-sur Yvette, p.139
\par\noindent
 Thuan T.X. et al. 1995, Ap.J., 445, 108 
\par\noindent
Weaver R., McCray R., Castor J., Shapiro P., Moore R., 1977, ApJ, 218, 377
\par\noindent
Woosley S.E., 1987, in {\em Nucleosynthesis and Chemical Evolution}
16th Saas-Fee Advanced Course, Geneva Observatory, p.1
\par\noindent
 Woosley S.E e Weaver T.A. 1995, Ap.J. Suppl., 101, 181 
\par\noindent

\end{document}